# A Wavelength Broker for Markets of Competing Optical Transport Networks

Abdulsalam Yassine
Alcatel-Lucent Inc.
Wireless Research and Development
Ottawa, Canada
e-mail: abdulsalam.yassine@alcatel-lucent.com

*Abstract*—The current trend in optical networks is to open the entire wholesale market to competition. As a result, we will see, instead of a single big market player, optical transport networks competing with each other to attract customer demand. This paper presents a wavelength broker who acts on behalf of enterprises, web host companies, financial firm etc. to buy certain number of wavelengths from such market. We present the system model, the interaction protocol and provide analysis of the competition. The simulation results of a business scenario are also recorded in the paper.

I. INTRODUCTION

The research on competition between optical transport service providers becomes a very important issue especially in the light of worldwide globalization of trade and telecommunications. Our study aims at integrating business models in the optical networks to increase the optical transport service provider's competitiveness and profitability in the telecommunication market. The history of telecommunication industry has a distinct pattern of transformation, starting as unregulated monopoly, later to a fierce competition, then to regulated monopoly, and most recently to (de)regulated competition [1]. The current trend in optical networks is to open the entire wholesale market to competition [2]. As a result, we will see, instead of a single big market player, several service providers competing with each other to attract customer demand. In this respect, our study provides the fundamental microeconomics aspect of the competition and the possibilities of business opportunities. Furthermore, it guides optical transport service providers through their strategic and long term network planning based on profit/loss analysis. The scenario of competitive pricing for flow assignment in multi-wavelength optical networks considers the following:
(a) Currently, the bandwidth in the optical networks is offered / provided in the form of wavelengths wholesale to carriers companies, enterprise organizations, web hosting firms, governments, etc. [3] [4]. The bandwidth of an optical light-path in the market over optical fiber is 2.5 Gbps and higher [5].
(b) Backbone long-haul optical network transport service providers (TSPs) are offering their services in terms of bandwidth measured in number of wavelengths.
(c) Customers such as Internet Service Providers (ISPs), enterprise organizations, financial firms, Application Service Providers (ASPs), governments, etc.) request an end-to-end connectivity (virtual links), which forms a virtual network topology. The links of this virtual topology have to be mapped on the underlying optical networks.

Plenty of research studies were conducted in the past, concerning competitive routing in communication networks [6], [7], [8] [9] from the microeconomic point of view, where users act in a selfish manner and compete for network resources until they reach an equilibrium. These studies were performed under the assumption that the user competes for scarce network resources in one network.

The novelty in our approach is the assumption that competitors now are the optical networks. They compete to allocate customers demand measured in number of wavelengths. The competition is assumed to be non-cooperative, which means, the competitors do not negotiate among each other. To the author's best knowledge, this approach has not been considered in the open scientific literature by anybody yet.

The rest of the paper is organized as follow: In the next section we describe the system model. In section 3 we provide detail analysis of the non-cooperative game of price competition. In section 4, we present routing an wavelength allocation followed by the simulation results in section 5. Finally, in section 6 we conclude the paper and provide directions for future work.

II. SYSTEM MODEL

*A. Opical wavelengh broker*

On figure 1 we present a typical scenario for network competition. We can identify a broker (broker acts on behalf of ISPs, web host company, financial firm, etc.) who sends a request to the optical transport service providers to buy certain number of wavelengths. In this model, the broker, representing many customers, decides the number of wavelengths to buy based on the cheapest price offer from the suppliers (optical network service providers) and the price-demand relationship $D(p)$. This price demand relationship is determined by the broker based on a pool of individual customers he represents.

The competition is initiated after the broker sends a request for logical link(s) connectivity (understood as virtual topology to be mapped on the physical networks) to the service providers. The service providers reply with price bids per unit wavelength. The broker replies to the service providers asking them to lower the bidding below the current minimum price without revealing the identity of the service provider who offered the minimum price. The interaction and the feedback between the broker and the optical transport providers continue until the end of the competition, when only one network can offer the lowest price. The broker either accepts the lowest price offered, or rejects it, if he cannot afford.

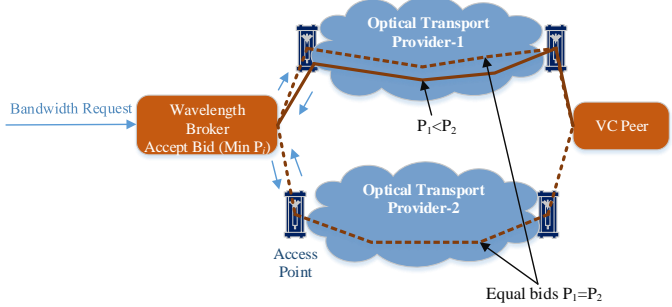

Fig. 1. Example of a broker and optical netwokr competition

### B. High-level Communication protocol

We develop a high-level communication protocol that supports an exchange of messages for competitive price setting based on the bids sent by the optical transport service providers and received by the broker. Before describing the protocol, we wish to point out the assumptions of the non-cooperative competition among suppliers.
(a) The broker sends end-to-end connectivity requests to the networks asking for a price per unit wavelength.
(b) Each network first replies with reasonable price based on market awareness. Each network is selfish and tries to increase its profit regardless of other networks.
(c) After the initial bidding, the broker submits to the networks the minimum price offer without revealing the identity of the service provider who offered this price. The broker asks the networks to go lower than the current minimum price. If the minimum price is the same as the price previously offered, then the network does not have to go lower. In this case, the service provider can wait for another round.
(d) If more than one service provider was able to answer with a lower price, then step c is repeated.
(e) The competition ends when only one service provider was able to offer a lower price.

The messages exchange between brokers and service providers are shown in table 1. The requested service is the virtual connection between nodes X and Y. The networks compete by offering price $P$ to win customer demand of wavelengths $D$. In figure 2, without loss of generality and for the sake of simplifying the protocol signaling, we will consider the case of two competing networks; Network A and Network B. It assumed that the duration of the customer connectivity request can take hours, days or weeks. The price competition is assumed to take milliseconds. Thus, the duration of the price competition is instantaneous compared to the duration of the request which might be hours, weeks or months. We present the sequence diagram of a general case of competition in which the broker accepts the final price (the price after finishing the competition) and the winner network can accommodate the requested number of wavelengths.

TABLE I.  MESSAGES OF COMMUNICATION PROTOCOL

| Direction | Messages | Comments |
|---|---|---|
| *Customer to Supplier (Broker to Network Service provider)* | *Reqc (X,Y)* | Initiate competition for the virtual link *X-Y* |
| | *ocl (X,Y,P)* | Offer a lower price than the current bid *P* for the virtual link *X,Y* |
| | *Nack (X,Y)* | The link request X-Y not granted |
| | *Ack (X,Y,D)* | The virtual link request X-Y is granted with D wavelengths to be accommodated |
| *Supplier to Customer (Network Service Provider to Broker)* | *offp(P,X,Y)* | The wavelength price offered for the virtual link X-Y is P dollars |
| *Exceptions* | *exc_1(D,P,X,Y) (supplier to customer)* | The requested demand D over virtual link X-Y exceeds the available network capacity or No capacity to accommodate even a single wavelength of the link X-Y |
| | *exc_2(X,Y,P) (customer to supplier)* | The offered price results in 0 wavelengths demand from the customer, the customer can not afford |

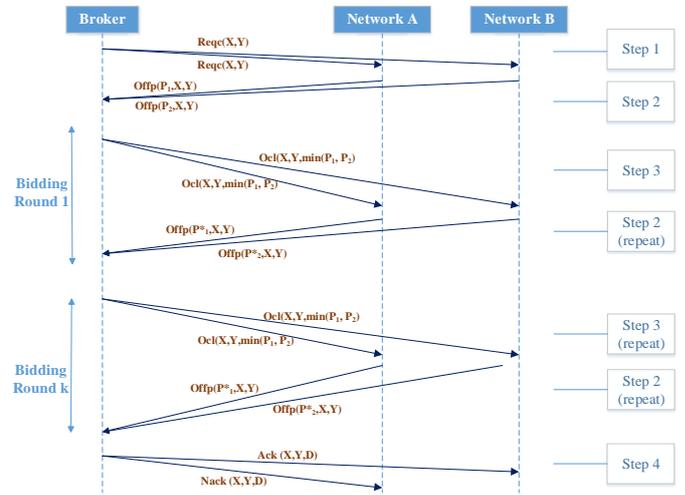

Fig. 2.  Interaction between the broker and optical transport networks

From figure 2, we can identify the following steps:
Step 1: The broker initiates the competition by sending his connection request to network A and network B.
Step 2: The suppliers (optical network service providers) respond with their price offerings.
Step 3: The broker responds to the offerings:
- The broker sends a request to the suppliers (optical networks) to undercut the minimum price of the offerings $P=min\ (P_1,\ P_2,...)$ without revealing the identity of the network that offers the minimum price and then the suppliers repeat step 2.
- If only one competitor was able to lower the price while the other competitor cannot, then this competitor is a winner and the price is final.

Step 4: the broker notifies the winner competitor and sends his demand, measured in number of wavelengths.

*C. Total and marginal cost in optical networks*

In this subsection, we introduce two terms from microeconomics, which play a pivotal role in the price competition and are necessary for our analysis: the Marginal Cost and the Total Cost. Based on these two terms we can determine the competitiveness of the optical network.

- The *Total production cost (TC)* expresses the total expenses required for producing $q$ items of a product and it is denoted as *TC (q)*.
- The *Production Marginal Cost (MC)* is defined as the production cost per unit output. It is defined as the change in the total production cost *(TC)* when producing an additional unit, i.e. $MC(q) = TC(q+1) - TC(q)$.

In the business case of wavelength allocation in optical networks the above terms are interpreted as follows:

- The production units are the wavelengths to be allocated.
- The unit production cost is the cost for allocating a single wavelength on a physical link. The cost per wavelength is determined by the need for wavelength regeneration in the optical switches and the optical fibers themselves. The power necessary for regenerating a wavelength can be as high as thousands of Watts [17][18]. This is the main reason why the total cost of using a fiber link depends on the flow (measured in number of wavelengths) over the physical link.
- The total cost is the sum of all physical link costs along the route for accommodating the requested number of wavelengths of the virtual link (point-to-point) connection.
- The marginal cost is the allocation cost of a single wavelength over a virtual link on the associated physical route of the network topology, given that the network has already allocated a number of wavelengths in the same virtual channel. In other words, it is the cost associated with one additional unit of production (known here as unit wavelength).

**Example**: Consider the network presented in figure 3 with the following assumptions:
1. Customer sends a connection request for a virtual channel from Vancouver to Toronto
2. We consider the following two alternative routes from Vancouver to Toronto:
   a. A-B and C-D-E-F-G
3. The capacities and prices of the physical links along these routes are given in figure 3.

The cheapest route of connection between Vancouver and Toronto is A-B. We assume that the allocation cost of an additional wavelength on a fiber is constant, as long the capacity of the fiber is not exceeded. It can be easily seen that for up to 8 requested wavelengths the whole demand can be accommodated on route A-B, because the capacity of the links along the route is not exceeded. For requests of 9 wavelengths and up, the additional wavelengths have to be allocated on the next cheapest route C-D-E-F-G. Requests over 17 wavelengths and more cannot be accommodated based on the fact that the capacity along these two routes is 8 wavelengths. From this example we can conclude that:

- The total cost *(TC)* is monotonically increasing and piece-wise linear.
- The marginal cost *(MC)* is monotonically increasing and piece-wise constant. It is based on the fact that the allocation cost of a single wavelength along the route is constant, as long the capacity of all links along the same route is not exceeded by the current accommodated requests.

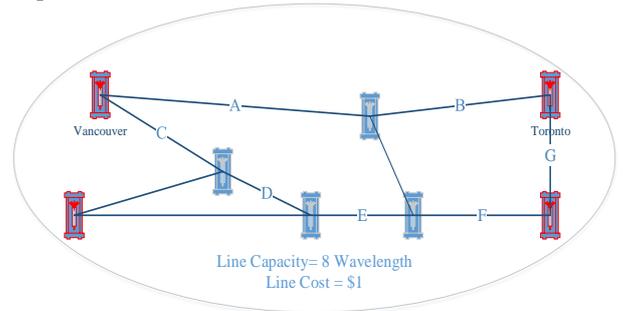

Fig.3. Example of an optical network

III. NON-COOPERATIVE GAME OF DYNAMIC PRICE COMPETITION

In this section we will consider a dynamic price competition game among non-cooperative optical networks. Non-cooperative means that the networks do not exchange information about their state topology, capacity, etc. In order to specify the process of competition we need to define it in terms of game theory. The game between the network service providers for attracting customer demand can be seen as a Non-cooperative game. The *competition* is offering competitive prices for allocating the requested wavelength demand. The *game strategy* is undercutting prices. *The players* are the networks. The *payoff* is the profit from allocating the requested demand. Each network is selfish and wants to maximize its profit independently and regardless from other networks. Next we explain the price undercutting game.

*A. Price undercutting game*

In price competition, we assume that the networks are driven by the awareness of each supplier regarding the competition [10]. The price competition is determined by the marginal cost *(MC)* of unit wavelength. The network with lower marginal cost *(MC)* has a competitive advantage to win the competition by offering a cheaper price [10]. These results are driven entirely by the assumption of homogeneous products being offered, which are in our case the wavelengths. Let us first present the price competition algorithm:

- The broker initiates a competition for a virtual channel (end-to-end) connection by sending the request to the competing networks.
- The competing networks respond with their initial price offers **B= (B₁, B₂, .., B_N)**, where *N* is the number of competing networks.
- After the initial bidding the broker submits to the networks the price of the minimum offer **P= min (P₁, P₂, ..., P_N)**, without revealing the identity of the competitor who offered this price, and ask the competitors to undercut the current minimum price. $P_1, .. P_N$ are the prices offered by the competitors for a unit wavelength.
- The network that offered the lowest price has no incentive to undercut its own price immediately. It will let the round go in case nobody cuts lower.
- The competing network *i* undercuts the current bidding price *P* by an undercutting step value $U_i$ as long as the corresponding marginal cost $MC_i$ *(i=1..N)* is not reached.
- The undercutting step value $U_i$ defined by network *i* is assumed to be stochastic. Each time the competitor undercuts the price, and the new price is **P*_i = P - U_i**.

*B. Stochastic price undercutting and equilibrium*

Stochastic price undercutting is a price competition in which the price undercutting step $U_i$ is chosen randomly in the interval $[L_i^{min}, L_i^{max}]$, where $L_i^{min}$ and $L_i^{max}$ are respectively the minimum and maximum possible undercutting steps for network *i*. The undercutting step $U_i$ is a uniformly distributed random variable in the interval $[L_i^{min}, L_i^{max}]$. The process of stochastic price undercutting has a lower limit, which is the equilibrium price. The price of the competing networks depends on the marginal cost. The equilibrium prices and the profit of two competing networks A and B will be as follows:

- if $MC_A > MC_B$ then equilibrium price of network B is $p_B = MC_A \pm e$
- if $MC_A < MC_B$ then equilibrium price of network A is $p_A = MC_B \pm e$
- if $MC_A = MC_B = MC$ then price $p_B = p_A = MC \pm e$, the user will choose Network A or Network B with equal probability.

Where *e* is bounded by the following inequality:
$min(L_1^{min}, L_2^{min}) \leq e \leq max(L_1^{max}, L_2^{max})$ for stochastic price undercutting.

The outcome of the competition is decided by the marginal production cost *(MC)* of each competitor [10]. The proof of the equilibrium can be found in the standard texts of industrial organization such as [11].

## IV. ROUTING AND WAVELENGTH ALLOCATION

In order to offer bids with minimal cost, a natural choice for routing in competitive optical networks is the minimum cost routing [12]. Therefore, the corresponding flow allocation has to be the minimum cost allocation, known also as minimum cost Routing Wavelength Allocation (RWA). The network is modeled as a set of nodes, *Nodes= {1,2,..,N}* and a set of optical links, $L= \{l_{xy}\}$ where $l_{xy}$ denotes the bidirectional link from node *x* to node *y*. Every optical link $l_{xy}$ is associated with a utilization cost $p_{xy}$. The set of available wavelengths is $\Lambda = \{1,2,..,W\}$. The network traffic is given as a set of connections, which have bandwidth demand $d_k$ and is associated with wavelengths and a sequence of optical links. Based on this, we can now designate the network parameters and variables, used for the routing and wavelength allocation.

The network parameters are:
$v_k$: the logical connection we want to realize
$d_k$: number of wavelengths demanded by customers for every logical channel
$f_{xy}$: the maximum number of wavelengths for the link $l_{xy}$
$p_{xy}$: the cost for allocation of a wavelength for the link $l_{xy}$

The network variables are:
$b^k_{w,xy}$: a flow binary variable, equal to one if a logical connection $v_k$ is carried on link $l_{xy}$ over fiber using wavelength *w*, and zero otherwise
$\Omega^k_w$: a binary variable, equal to one if connection $v_k$ is carried on wavelength *w*, and zero otherwise

The production cost optimization function of wavelength allocation is as follows:

$$\min c(y) = \sum_{k=1}^{k} \sum_{w=1}^{W} \sum_{l_{xy} \in L} p_{xy} \cdot b^k_{w,xy}$$

subjected to constraints

- Flow conservation constraint: for every node x and neighboring nodes *j*

$$\sum_{j \neq x} b^k_{w,xj} - \sum_{j \neq x} b^k_{w,jx} = \begin{cases} +\Omega^k_w & \text{if } x \text{ is source of } v_k \\ -\Omega^k_w & \text{if } x \text{ is destination of } v_k \\ 0 & \text{otherwise} \end{cases} \quad (1)$$

Equation (1) states that a connection $v_k$ entering node x on wavelength *w* must leave the node on the same wavelength, thus ensuring wavelength continuity.

- Capacity constraint

$$\sum_{w=1}^{W} \sum_{k=1}^{K} b^k_{w,xy} + \sum_{w=1}^{W} \sum_{k=1}^{K} b^k_{w,yx} \leq f_{xy} \quad (2)$$

Equation (2) specifies the capacity limit of every optical link, where $f_{xy}$ is the maximum number of wavelengths for link $l_{xy}$. The wavelength can be directed on the link from node *x* to node *y* or from node y to node x, but are not bi-directional simultaneously.

- Constraint for one traffic direction over single wavelength

$$b^k_{w,xy} + b^j_{w,yx} \leq 1 \quad (3)$$

Equation (3) specifies that while the links (i.e. one fiber) are bi-directional on a single wavelength, the communication is only in one direction. As indicated earlier, $b^k_{w,xy}$ is a flow variable equal to one if a logical connection $v_k$ is carried on link $l_{xy}$ over fiber using wavelength *w*, and zero otherwise,

$$b^k_{w,xy} \in \{0,1\}$$

- Traffic demand constraint

$$\sum_{w=1}^{W} \Omega_w^k = d_k \text{ , where } \Omega_w^k \in \{0,1\} \qquad (4)$$

Equation (4) ensures that the requested demand, interpreted as number of wavelengths for every optical connection, is actually allocated throughout the network.

This formulation can be cast as Mixed Integer Programming Formulation (MIP) [13]. Next, we present a case study on how a transport provider can utilize the existing resources of a network; improve the network planning with regard to profitability and survivability.

## V. SIMULATION RESULTS

In this section, we describe the model scenario used for the simulations. In figure 4, we consider two service providers competing with each other to allocate the brokers' requests for end-to-end connectivity on virtual channels VC1, VC2, and VC3. We wish to point out that the most widely used network in the open literature is the National Science Foundation (NSF) network [14], which has 16 nodes. Some researchers as in [15], simulate their results using network size of 6 nodes, others like in [16] use a network size of 8 nodes. In our simulations we used the AMPL modeling language software with direct link to the solver CPLEX 7.0, this software is limited to 300 variables, this is the reason we have only 8 nodes in the network.

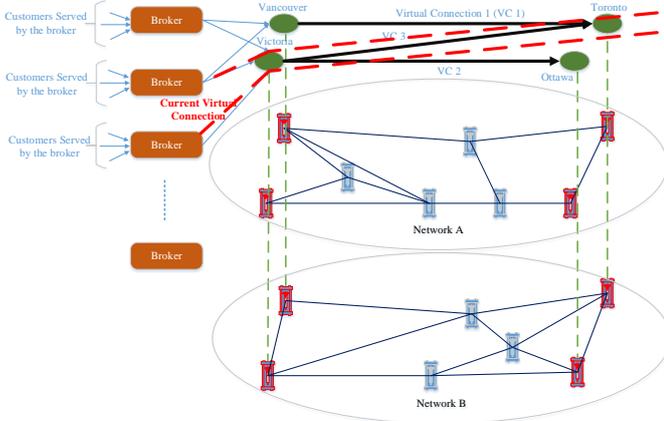

Fig.4. Optical network competition scenario and virtual connection allocation

Our objective is to determine the profitability of the optical network when offering prices based on competition between the transport service providers. Furthermore, we need to provide a recommendation on network upgrade based on profit/loss analysis of the wavelengths sale over the virtual channels. In order to do this, we first need to determine the operational profit and the amount of wavelengths sale on virtual channels VC1, VC2, and VC3 for both networks. Then we provide the analysis based on these results.

Figure 5 shows the number of active connections requests on each virtual channel. We assume that brokers have different price-demand functions for each virtual channel. For example the price-demand function on virtual channel 1 and 2 could be linear while the price demand function on virtual channel 3 is non-linear. The allocation of the wavelength requests is performed using the min-cost optimization routing as described in section IV. In figure 6, we show the number of active requests served on network A and network B. In figure 7, we show the resulting profit of Network A and Network B after the competition. Finally, in figure 8 and 9 we show the profit percentage form wavelength sale on each virtual channel for network A and network B.

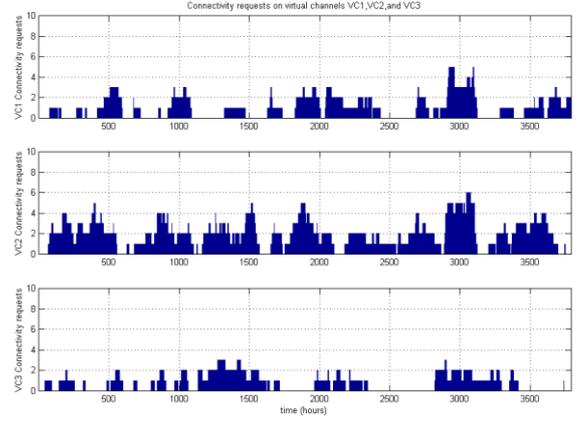

Fig. 5. Number of connection requests on VC1, VC2 and VC3

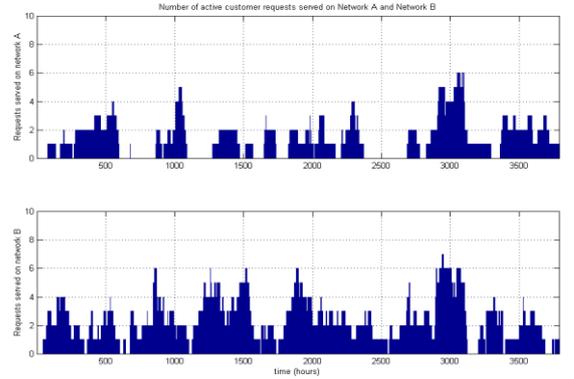

Fig.6. Number of active requests attracted and served by networks A and B

As we can see from figure 7, the performance of Network A in terms of profitability is much less when compared to Network B. The average profit of network A for 50 simulation runs is $21962, while the average profit of network B is $187506. Figure 8 and figure 9, show the resulting profit percentage of Network A and Network B from wavelengths sale. The profit from wavelengths sale as shown in figures 8 and 9 indicates that Network B was better off when competing for customer requests on virtual channel VC3. In order to boost the competitive advantage of Network A, a physical connectivity upgrade along the routes that serve virtual channel VC3 needs to be considered. The decision where to place additional nodes and the corresponding appropriate links connections is a wide area of study, which is beyond the scope of this paper. However, in this study, the profit/loss analysis can be used as a criterion, which helps network designers and network architectures in the upgrading decision of the optical network to increase its profitability.

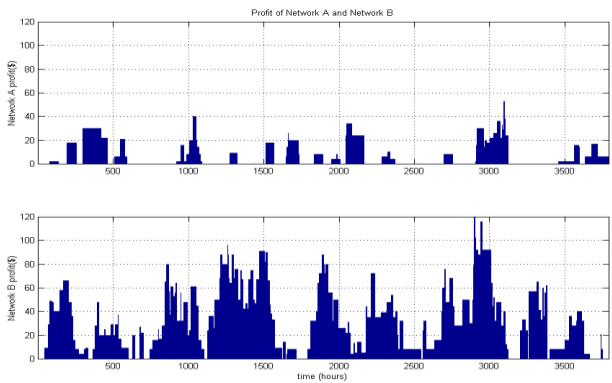

Fig.7. Resulting profit of operating the networks

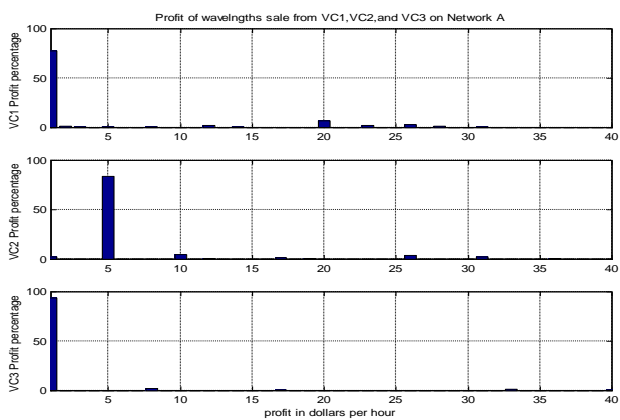

Fig.8. Profit percentage from wavelengths sale on Network A

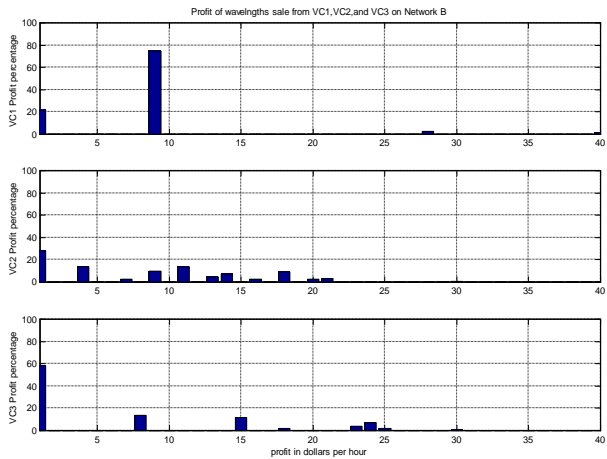

Fig. 9. Profit percentage from wavelengths sale on Network B.

## VI. CONCLUSION AND FUTURE WORK

This paper is based on an economic feasibility study for a service provider, offering transport network services in a competitive market where a pricing decision has to be chosen carefully to acquire market shares and stay profitable. We have proposed a microeconomic approach for routing and wavelength allocation in competitive non-cooperative disjoint optical networks. We modeled the pricing decisions as a game between two players and developed a messaging protocol supporting the competition process. Our plan for the future is to study other scenarios of competition and price undercutting. Also the implementation of heuristic (sub-optimal) routing strategies in the optical networks and their effects on the network competitiveness and profitability.